\documentclass[pra,aps,preprint,showpacs]{revtex4}

\usepackage{amssymb}
\usepackage{epsfig}
\usepackage{graphicx}

\begin{document}
\title{Casimir repulsion and metamaterials}
\author{I. G. Pirozhenko\footnote{Permanent address: Bogoliubov Laboratory of Theoretical
Physics, JINR, 141980 Dubna, Russia},
A. Lambrecht}
\address{Laboratoire Kastler Brossel, CNRS, ENS,
UPMC - Campus Jussieu case 74, 75252 Paris, France}
\email{Irina.Pirozhenko@spectro.jussieu.fr,  Astrid.Lambrecht@spectro.jussieu.fr}
\begin{abstract}
We analyze the conditions for getting the Casimir repulsion between two nonequal plates.
The force between plates with magnetic permeability defined by Drude or Lorentz models
is calculated. The  short and long distance limits of the force are derived.
The Casimir set-up with the hypothetical perfect matching metamaterial is discussed.
We put into question the possibility of getting repulsion within the design of
metamaterials based on metallic inclusions.
\end{abstract}
\maketitle

\section{Introduction}
The study of Casimir repulsion which always presented  purely
academic interest has moved in the last years to a more practical
field. It is inspired by new precision measurements of attractive
Casimir force and the development of micro(nano)
electromechanical machines where the Casimir repulsive force, if
any, might resolve the stiction problem~(see \cite{Iannuzzi} and ref. therein).

In the present paper we address the Casimir repulsion between
parallel plates owing to the optical properties of the material.
From Lifshtz formula~\cite{Lifshitz} it follows that the force
between two parallel plates
is repulsive if  the gap
between them is filled up with a material that satisfies the
inequality $\varepsilon_1({\rm i}\xi)<\varepsilon_3({\rm i}\xi)<\varepsilon_2({\rm i}\xi)$~\cite{Israelachvili},
where  $\varepsilon_1$ and $\varepsilon_2$  stand for the dielectric permittivities of the plates,
and $\varepsilon_3$ corresponds  to that of the filling~\cite{Iannuzzi, Henkel}. We do not consider this set-up
in the present paper.

The force may become repulsive if one of the plates has nontrivial magnetic permeability, $\mu\ne1$.
This possibility was not seriously regarded since
for "natural" materials, where the magnetization of the
system is due to the movement of the electrons in the atoms,
$\mu(\omega)=1$ at visible range~\cite{Landafshits}.
However in composite materials if the inclusions are smaller than the
wavelength, but larger than the atomic size the effective dielectric and magnetic functions
can be introduced as a result of local field averaging.
That is why  the artificial materials~\cite{IEEE_pendry} with magnetic response arising from
micro (nano) inclusions have recently become good candidates for observing the Casimir repulsion.

The Casimir repulsion for materials with dielectric permittivity and magnetic permeability
that do not depend on the frequency was considered in~\cite{Klich}.
The case of Drude-Lorentz dispersion relations for
dielectric permittivity and magnetic permeability of the metamaterials was analyzed
in~\cite{Henkel}. The upper limits for Casimir attractive and
repulsive forces between multilayered structures at finite temperature were
established. Starting from the Lifshitz formula it was shown that at zero temperature
 $-7/8 \,F_C(L) \le F(L) \le F_C(L)$, where $F_C$ is the force between perfect conductors.

In~\cite{Leonhardt} the Casimir repulsion due to the presence of dispersive anisotropic
materials with gain was first predicted. These media are beyond the scope of our paper.

In Section II we start from the basic formulas for the Casimir force between two nonequal
plates and give the conditions for the Casimir repulsion. We consider several models for
the plates and  get analytic results for the force at short and long distances. In
section III the perfect matching metamaterial is discussed.  We also consider recently
reported optical range metamaterial as a candidate for Casimir repulsion. In Conclusion
we discuss the possibility of getting repulsion within the design of metamaterials based
on metallic inclusions.

\section{The Casimir force between non-equal mirrors}
The Casimir force between two flat mirrors separated by  a
distance $L$ is given by
\begin{equation}
F(L)=-\frac{\hbar}{4\pi c^4}\sum\limits_{\rho}\int\limits_0^{\infty} d\omega \,\omega^3
\int\limits_1^{\infty} d\alpha \,\alpha^2 \,\frac{r^{\rho}_A\,
r^{\rho}_B}{e^{2\alpha\omega L/c}-r^{\rho}_A\, r^{\rho}_B}.  \label{eq1}
\end{equation}
Here $r^{\rho}({\rm i}\omega,\alpha)$, $\rho=TE, TM$,  are the reflection
coefficients at imaginary frequencies for the mirrors facing vacuum
\begin{eqnarray}
r^{TM}=\frac{\sqrt{(n^2-1)+\alpha^2}- \varepsilon
\alpha}{\sqrt{(n^2-1)+\alpha^2}+\varepsilon \alpha},\;\;
r^{TE}=-\frac{\sqrt{(n^2-1)+\alpha^2}- \mu
\alpha}{\sqrt{(n^2-1)+\alpha^2}+\mu \alpha}
\label{eq2}
\end{eqnarray}
with  $n=\sqrt{\varepsilon \mu}$, $\varepsilon=\varepsilon({\rm i}\omega)\geq1$,
$\mu=\mu({\rm i}\omega)\geq1$ ~\cite{Landafshits,MostKlim}.  The variable $\alpha$ is the ratio
of the transversal wave-vector at imaginary frequency $\kappa=\sqrt{\omega^2/c^2+k^2}$ to
the inverse wave-length $\omega/c$.

The sign of the force is defined by the sign of the integrand in~(\ref{eq1}). As
$|r({\rm i}\omega,\alpha)|\leq 1$, a "mode" $\{\omega,\alpha\}$ gives a repulsive contribution
to the force if the corresponding reflection coefficients of the mirrors $A$ and $B$ have
opposite signs. This happens if  the mirrors are different, $r_A \neq r_B$, and  at least
one mirror has nontrivial magnetic permeability. In~\cite{Henkel, Genet} it was proved
that no multilayered dielectric mirror can reverse the sign of the force.

At the lower limit of the integral over $\alpha$ the reflection
coefficients for TE and TM modes coincide
$$
\lim_{\alpha \to1} r^{TM}=\lim_{\alpha\to1}r^{TE}=
(\sqrt{\mu}-\sqrt{\varepsilon})/(\sqrt{\mu}+\sqrt{\varepsilon}).$$
At the upper limit the reflection coefficients for TE and TM modes
are different:
$$
\lim_{\alpha
\to\infty}r^{TM}=(1-\varepsilon)/(1+\varepsilon)\leq0, \quad
\lim_{\alpha\to\infty}r^{TE}=-(1-\mu)/(1+\mu)\geq0.$$
For fixed $\omega$ one can find the values of $\alpha$ where the reflection coefficients change their signs:
$
\alpha_0^{TM}(\omega)=\sqrt{n^2-1}/\sqrt{\varepsilon^2-1}$, \
$\alpha_0^{TE}(\omega)=\sqrt{n^2-1}/\sqrt{\mu^2-1}$.
We split the integral over $\alpha$ in (\ref{eq1})  in two:  $F_1^{\rho}(L)$ and $F_2^{\rho}(L)$.
In the former the integration goes from $1$ to $\alpha_0^{\rho}$, and in the latter from $\alpha_0^{\rho}$
to $\infty$.

The reflection coefficients have no extremum with respect to
$\alpha$. If $\mu=1$,  $\lim_{\alpha\to\infty}r^{TE}=0$, then $r^{TE},r^{TM}\le0$ at all frequencies.
When $\mu<\varepsilon$, $r^{TM}$ is always negative,
falling from $r^{TM}(\alpha=1)$ to $r^{TM}(\alpha=\infty)$, while
$r^{TE}$ is monotonously growing from negative to positive values.
If $\mu>\varepsilon$, $r^{TE}$
remains positive and  growing with $\alpha$,  as $r^{TM}$
decreases  from positive values at $\alpha=1$ to negative at
$\alpha=\infty$.
Below we analyze the force in these situations considering Drude or Lorentz models
for the mirrors.
\subsection{Two  non-magnetic mirrors.}
When $\mu=1$, neither $r^{TM}$ nor $r^{TE}$ changes the sign in the course of the
integration over $\alpha$ from 1 to $\infty$. Hence if both mirrors are non-magnetic the
function~(\ref{eq1}) is always positive, and the force is attractive at all distances. As
$|r^{TM}|\geq|r^{TE}|$, the contribution of the TM modes to the Casimir force~(\ref{eq1})
exceeds the TE contribution. For the mirrors described by the Lorentz model
$\varepsilon_i(\omega)=1-\omega_{e,i}^2/(\omega^2-\omega_{0}^2+{\rm
i}\gamma_{e,i}\omega)$, $i=A,B$, at short distances $L<< \lambda_{ei}$,
$\lambda_{ei}=2\pi c/\omega_{ei}$, with $\omega_{eA}<\omega_{eB}$ we get
\begin{equation}
F\simeq -\frac{\hbar}{8 \pi^2 L^3}
\frac{\Omega_{2A}}{2}\sum_{k=0}^{\infty}\,G_k\,\left(1-
\frac{\Omega_{2A}^2}{\Omega_{2B}^2}\right)^k, \label{eq3}
\end{equation}
where $\Omega_{1i}^2=\omega_{ei}^2/2$,
 $\Omega_{2i}^2=\omega_{ei}^2/2+\omega_{0}^2$, $i=A,B$, and
 $G_k=G_k(\omega_{0}/\omega_{eA}, \omega_{0}/\omega_{eB})$.
The absorbtion in the material influences more the small
frequencies  which make the decisive contribution to the force at large plate separations.
That is why the relaxation parameters $\gamma_{e,i}$ do not enter  the short
distance asymptote.

For the mirrors described by Drude model,  $\omega_0=0$, the result is simplified to
\begin{equation}
F\simeq -\frac{\sqrt{2}}{32}\frac{\hbar \omega_{eA}}{\pi^2 L^3}\sum_{k=0}^{\infty}\,G_k\,\left(1-
\frac{\omega_{eA}^2}{\omega_{eB}^2}\right)^k,
\label{eq5}
\end{equation}
with $G_0\simeq1.744,\;G_1\simeq0.436,\;G_2\simeq0.215,\;G_3\simeq0.133,...$

The long distance limit is obtained by expanding the integrand
in~(\ref{eq1}) in powers of the small parameter $\lambda_{eA}/L$ or
$\lambda_{eB}/L$ and given by
$$
F|_{L>>\lambda_{eA},\,\lambda_{eB}}=\eta \,F_{Cas}(L) ,\quad
\eta\approx1-4 \,(\lambda_{eA}+\lambda_{eB})/(3\pi L).$$

\subsection{Mirror A is purely dielectric, mirror B is purely
magnetic.} Let  mirror A be purely dielectric, $\mu_A=1$.  Then
$r_A^{TE},r_A^{TM}\leq0$. The mirror B is purely magnetic,
$\varepsilon_B=1$, $r_B^{TE},r_B^{TM}\geq0$. Then both TE and TM modes are repulsive at all
frequencies,  and the force {\it is repulsive at all distances}.

The short distance limit for plasma models,
$\varepsilon_A=1+\omega_{eA}^2/\omega^2,\;\mu_B=1+\omega_{mB}^2/\omega^2$,
is given by
\begin{equation}
F(L)\approx \frac{ \sqrt{2}}{64}\, \frac{\hbar}{\pi c^2}\,
\frac{(\omega^2_{eA}\,\omega_{mB}+\omega_{mB}^2 \,\omega_{eA})}{L}. \label{no-pl}
\end{equation}
The short distance attraction due to the interaction between surface plasmons is absent
in the present case. The TM-plasmonic mode of purely dielectric mirror and TE-plasmonic
mode of purely magnetic mirror are not coupled, and therefore do not contribute to the
Casimir force. It results in the unusual short distance asymptote~(\ref{no-pl}).

At long distances we get
$$F|_{L>>\lambda_{eA},\,\lambda_{eB}}=\eta \,F_{Cas}(L), \quad
\eta\approx-7/8+ 7 \,(\lambda_{eA}+\lambda_{eB})/(6\pi L).$$
The repulsive force coinciding with the first term of this expansion was obtained by Boyer~\cite{Boyer}
for two non-dispersive mirrors with $\varepsilon_A=\infty$, $\mu_A=1$ and $\varepsilon_B=1$, $\mu_B=\infty$.

\subsection{Mirror A is purely dielectric, mirror B is mainly
dielectric} Let the mirror A be purely dielectric, $\mu_A=1$, with
$r_A^{TE},r_A^{TM}\leq0$,  and  mirror B  mainly dielectric, so
that $1\leq\mu_B\leq\varepsilon_{B}$ for all frequencies. Then
$r_B^{TM}\leq0$, and the contribution of the TM modes is attractive
at all distances,  whereas $r_B^{TE}\leq0$  only for
$\alpha<\alpha_0^{TE}$ corresponding to  mirror B. When  $\alpha>\alpha_0^{TE}$ the signs of
the TE reflection coefficients for the mirrors A and B are opposite.
The sign of the force is the result of the balance between $F^{TM}(L)\le0$, $F^{TE}_1(L)\le0$, and
$F^{TE}_2(L)>0$.

If $\mu_B(\omega)=\varepsilon_{B}(\omega)$, then $\alpha_0^{TE}=1$,
and the contribution of TE modes is entirely positive,
$F^{TE}(L)\ge0$.  However as  $r_B^{TM}=-r_B^{TE}\le0$ and
$|r_A^{TE}({\rm i}\omega)|<|r_A^{TE}({\rm i}\omega)|$, the total force
is attractive (negative). Consequently, when mirror B is mainly dielectric,
$\mu_B(\omega)<\varepsilon_{B}(\omega)$, the force is attractive at
all distances as well.
At short distances it is determined by the modes of TM polarization
and given by (\ref{eq3}).


\subsection{Mirror A is purely dielectric, mirror B is mainly
magnetic.} Let the mirror A be purely dielectric, $\mu_A=1$, with
$r_A^{TE},r_A^{TM}\leq0$,  and  mirror B is mainly magnetic, so that
$1\leq\varepsilon_{B}\leq\mu_B$ for all frequencies. Then
$r_B^{TE}\geq0$, and the contribution of the TE modes is repulsive
at all distances, whereas $r_B^{TM}\geq0$  only for
$\alpha<\alpha_0^{TM}$.  When $\alpha<\alpha_0^{TM}$ the signs of
the TM reflection coefficients of the mirrors A and B coincide. Thus $F^{TE}(L)\geq0$,
$F_1^{TM}(L)\geq0$, $F_2^{TM}(L)< 0$.
The negative term  $F_2^{TM}(L)$ becomes dominant for the TM modes
at  distances $L\leq c/(\omega_0^B \alpha_{0}^{TM})$. At short
distances the TM reflection coefficients are larger than the TE
ones. The total force is attractive for short plate separation and
repulsive at middle and long distances. For the short distance
asymptote  see Eq. (\ref{eq3}).

\begin{figure}[ttp]
\centerline{\epsfig{file=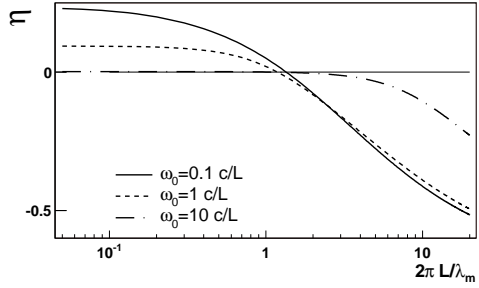,width=7cm}}
\caption{Reduction factor $\eta=F/F_C$ as a function of dimensionless
distance $\Lambda=2\pi L/\lambda_{m}=\omega_m L/c$}
\label{fig1}
\end{figure}

Fig.~\ref{fig1} gives the reduction factor of the force between purely dielectric mirror A
described by Drude model,
$
\varepsilon_A({\rm i}\omega)=1+\omega_p^2/[\omega(\gamma+\omega)]$, $\omega_p=10 c/L$,
$\gamma=0.01 c/L$,  $\mu_A=1$,
and  mirror B with
$
\varepsilon_B({\rm i}\omega)=1+\omega_e^2/(\omega^2+\omega_0^2+\gamma_e \omega)$,
$\mu_B({\rm i}\omega)=1+\omega_m^2/(\omega^2+\omega_0^2+\gamma_m \omega)$,
where
$\omega_e=\gamma_e=\gamma_m=c/L,\; \omega_0=0.1 c/L,\, 1 c/L,\, 10 c/L $.  Varying $\omega_m L/c$ one can see that
the curves cross the horizontal axis when  $\omega_m\sim\omega_e$, in other words, when the material
turns into mainly magnetic.  For a metamaterial with $\omega_m>\omega_e$,
$\omega_0/2\pi\sim 3\cdot10^4$ GHz the force becomes repulsive at the distances $L>10^{-6}$ m.

\section{The perfect matching metamaterial from the Casimir force viewpoint}
If the real part of the refractive index, $n=n'+ {\rm i}n''$, is negative, then the
transversal wave-vector, $k_z=(n^2 \omega^2/c^2 -k^2_{||})^{-1/2}$,  for the propagating
waves $k_z(n',n'')=-k_z(-n',n'')$, and for the evanescent waves
$k_z(n',n'')=k_z(-n',n'')$. It follows from the analysis of
$k_z=\sqrt{\rho_1\rho_2}\exp({\rm i}\frac{\phi_1+\phi_2}{2})$ with $n'<0$. Here
$$
\rho_1=\left(\left[n'
\omega/c-|k_{||}|\right]^2+\left[n''\omega/c\right]^2\right)^{1/2},\quad
\rho_2=\left(\left[n'\omega/c+|k_{||}|\right]^2+\left[n''\omega/c\right]^2\right)^{1/2},
$$
$$
\phi_{1}=\arctan\frac{n''}{-|n'|-|k_{||}| \frac{c}{\omega}}+ \pi, \quad
\phi_{2}=\arctan\frac{n''}{-|n'|+|k_{||}| \frac{c}{\omega}}+\left\{{\begin{array}{ll}0,&
\mbox{if}\;\;  n'\frac{\omega}{c}>|k_{||}| \\ \pi,&\mbox{if}\;\;
n'\frac{\omega}{c}<|k_{||}|
\end{array}}\right..
$$

When the dielectric permittivity, magnetic permeability and the  refractive index tend to
$-1$~\cite{Veselago,Pendry}, and the absorbtion in the material is negligible,  the ideal
situation of perfect matching between material and vacuum could be achieved. It means
that the transversal wavevector for the propagating waves $k_z\to-k_z^{vac}$.
Consequently, the reflection coefficients of the propagating waves vanish,
$r^{TM},r^{TE}\to0$, on the interface vacuum-metamaterial.

Let us consider two perfectly conducting mirrors one coated with a metamerial.
Equation~(\ref{eq1}) requires that $\varepsilon({\rm i}\omega)$ and $\mu({\rm i}\omega)$ are real
positive functions in accordance with causality~\cite{Landafshits,MostKlim}. Obviously,
this condition is not satisfied for the material with constant negative dielectric
permittivity and magnetic permeability. The straightforward substitution
$\mu=\varepsilon=-1$ in the formula~(\ref{eq1}) for the Casimir force leads to positive
force at $L>d$, and divergency at $L\leq d$. Moreover, the energy density of the
electromagnetic field inside the non-dispersive material with $\varepsilon=\mu=-1$ is
negative. Therefore its very existence contradicts the Pointing theorem,
$W=[\partial(\omega\varepsilon(\omega))/\partial\omega] E^2+
[\partial(\omega\mu(\omega))/\partial\omega]H^2 < 0$.  On the contrary, the real parts of
$\varepsilon(\omega)$ and $\mu(\omega)$ may tend to $-1$ at a certain frequency, leaving
the energy density of the electromagnetic field positive, $W\geq 0.$

\begin{figure}[ttp]
\epsfig{file=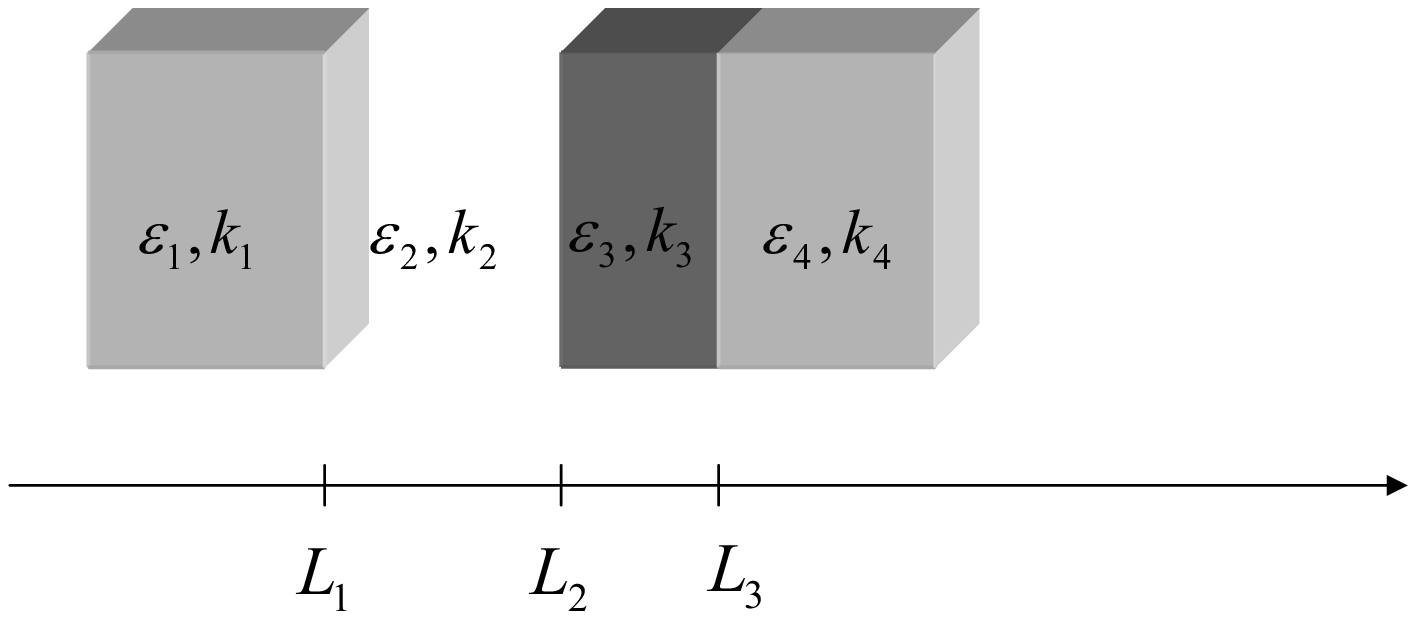,width=6.5cm}
\epsfig{file=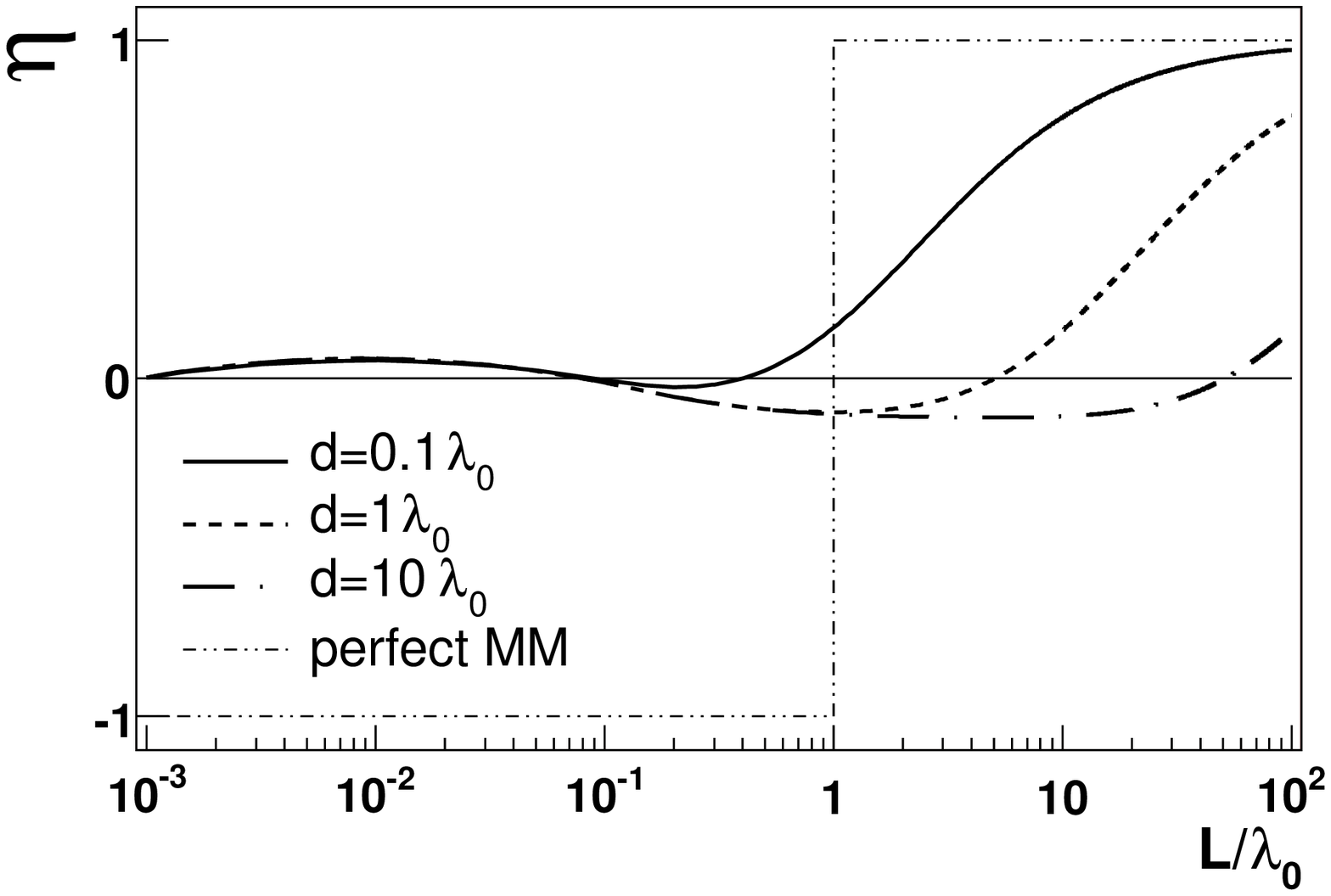,width=7cm}
\caption{The reduction factor for two mirrors, one coated by metamaterial; $d=L_3-L_2.$}
\label{fig8}
\end{figure}

The Casimir energy  for a multilayered system given on Fig.~\ref{fig8}
(left) can be defined in terms of the scattering phase
shift $\delta$~\cite{Lambrecht, bordag}
\begin{eqnarray}
\frac{E_C}{A}=\frac{\hbar}{2}\sum_{\rho}\int\frac{d^2k_{||}}{(2\pi)^2}
\left\{\sum\limits_{\sigma}\omega^{sp}_{\sigma}(k_{||})+
\int\limits_0^{\infty}\frac{dk_1}{\pi}\,\omega(k_{||},k_1)\frac{\partial \delta(k_1)}{\partial k_1}\right\}.
\label{eq22}
\end{eqnarray}
The first term in~(\ref{eq22}) corresponding to bound states (surface plasmons) is absent
in the case of perfect mirrors.

The Maxwell equations  are reduced to
$\phi''(z)-\{k^2_{||}-\varepsilon_i\mu_i\omega^2/c^2\}\phi(z) =0$, for TM modes, and
$\psi''(z)-\{k^2_{||}-\varepsilon_i\mu_i\omega^2/c^2\}\psi(z)
=0$, for TE ones, $i=1..4$.
To get the reflection and transmission coefficients and the phase shift one has to solve
the system of matching conditions, two for each interface in each polarization:
$\mu_+ \psi_+=\mu_-\psi_-$, \     $\psi'_+=\psi'_{-}$;
$\varepsilon_+ \phi_+=\varepsilon_-\phi_-$, \   $\phi'_+=\phi'_{-}$.
Here $\mu_{\pm}$, $\varepsilon_{\pm}$, $\phi_{\pm}$, $\psi_{\pm}$ stand
for the values of the functions when  $z$  tends to the interface
from the right (left).  Taking the solutions  for TE modes in the
form $\phi_i=A_i e^{{\rm i} k_i z}+B_i e^{-{\rm i} k_i z}$, $i=1..4$, $B_4=0$,
$k_i=(\varepsilon_i\mu_i\omega^2/c^2-k^2_{||})^{1/2}$,
and substituting them into the matching condition we arrive at the system for the coefficients $A_i$, $B_i$.
Then the transmission  and reflection coefficients are given by the ratios
 $t(k_1)=A_4/A_1$, $r(k_1)=B_1/A_1$. The TM transmission and reflection coefficients are obtained by replacing
  $\varepsilon_i\leftrightarrow\mu_i$.

With $k_1=k_4,\; k_2=-k_3, \; \varepsilon_2=\mu_2=\mu_1=1, \; \varepsilon_3=\mu_3=n_3=-1$  the
scattering phase shift of the the four layered system is reduced to
\begin{eqnarray}
2\,{\rm i}\,\delta(k_1, |a'|)=\ln\frac{t(k_1)}{t(-k_1)}=
\ln\frac{1-r^2 e^{-2{\rm i} k_2 |a'|}}{1-r^2\, e^{2{\rm i} k_2 |a'|}},
\label{eq23}
\end{eqnarray}
where  $a'\equiv 2 L_2-L_1-L_3\geq0$, \  $r^{TM}=(k_2 \varepsilon_1-k_1)/(k_2 \varepsilon_1+k_1)$, \
$r^{TE}=(k_2 -k_1)/(k_2+k_1)$.  Here the limit of infinite plate separation is subtracted.
If the thickness of the metamaterial is larger than the separation of the mirrors, $a'\equiv 2 L_2-L_1-L_3<0$,
then $\delta(k_2, -|a'|)=-\delta(k_2,|a'|)$.

When layers 1 and 3 are made of perfect metal, $r^{TE}=r^{TM}=1$, we arrive at the Casimir result
$$E(a')=\mp A\frac{\hbar c \pi^2}{720 |a'|^3} \Rightarrow F(a')=\mp A\frac{\hbar c \pi^2}{240 |a'|^4}$$
with upper (lower) sign corresponding to positive (negative) effective distance $a'$ and attractive
(repulsive) force. The force diverges at $a'=0$. This result can not be recovered as a limiting case of
any dispersive model of a metamaterial consistent with Kramers-Kronig relations.

The account for finite conductivity of the metal and dispersion in the metamaterial leads to a
finite result at a plate separation equal to the thickness of the MM-coating.
Fig.~\ref{fig8} (right)  gives the reduction factor  for the force between two equal mirrors one of which
is coated with
mainly magnetic metamaterial. The mirror A and the substrate of mirror B are described by  Drude  model.
The coating of mirror B is a metamaterial with effective permittivity and effective permeability given by
Drude-Lorentz model. All parameters are normalized on
the position of the resonance $\omega_0=2\pi\,c/\lambda_0$, \
$
\omega_p/\omega_0=10$, \ $\gamma/\omega_0=\gamma_e/\omega_0=\gamma_m/\omega_0=0.01$,
$\omega_e/\omega_0=2$, \ $\omega_m/\omega_0=3$.
Here $d=L_3-L_2$ is the thickness of the metamaterial. For comparison we plot the reduction
factor $\eta'=F(a')/F_C(a'),\, a'=(L/\lambda_0-1)$ for  two perfect
metals one coated by perfect matching metamaterial.

We see that if the mirror B is two-layered, the force
changes the sign two times.  For short and intermediate distances, $L\leq d$,  the contribution of the reflections from the
interface between
the layer of the metamaterial and the substrate is small, and the behavior  corresponding to bulk MM-mirror
is reproduced  (compare with Fig.~\ref{fig1}).
 At large distances, $L>d$,  the fluctuations "feel"
the presence of the substrate, and the reflection coefficients approach the ones for the substrate.
The force becomes attractive, achieving the values typical for metals.
When the thickness of the metamaterial is smaller than its characteristic wavelength $\lambda_0$,
the layer becomes transparent for fluctuations with $\omega>\omega_0$, and the region of repulsion is considerably
narrowed (solid curve in Fig.~\ref{fig8}).

Further we discuss recently reported low loss optical metamaterial~\cite{OPT_LETT_1}.
For describing the material we use the effective media
approach, considering anisotropic compound material as a homogeneous
media having  effective dielectric and magnetic functions.
To evaluate the appropriate  parameters of the effective media we took the
complex permittivity and permeability plots from Fig.3 in
\cite{OPT_LETT_1}.

\begin{figure}[tbp]
\epsfig{file=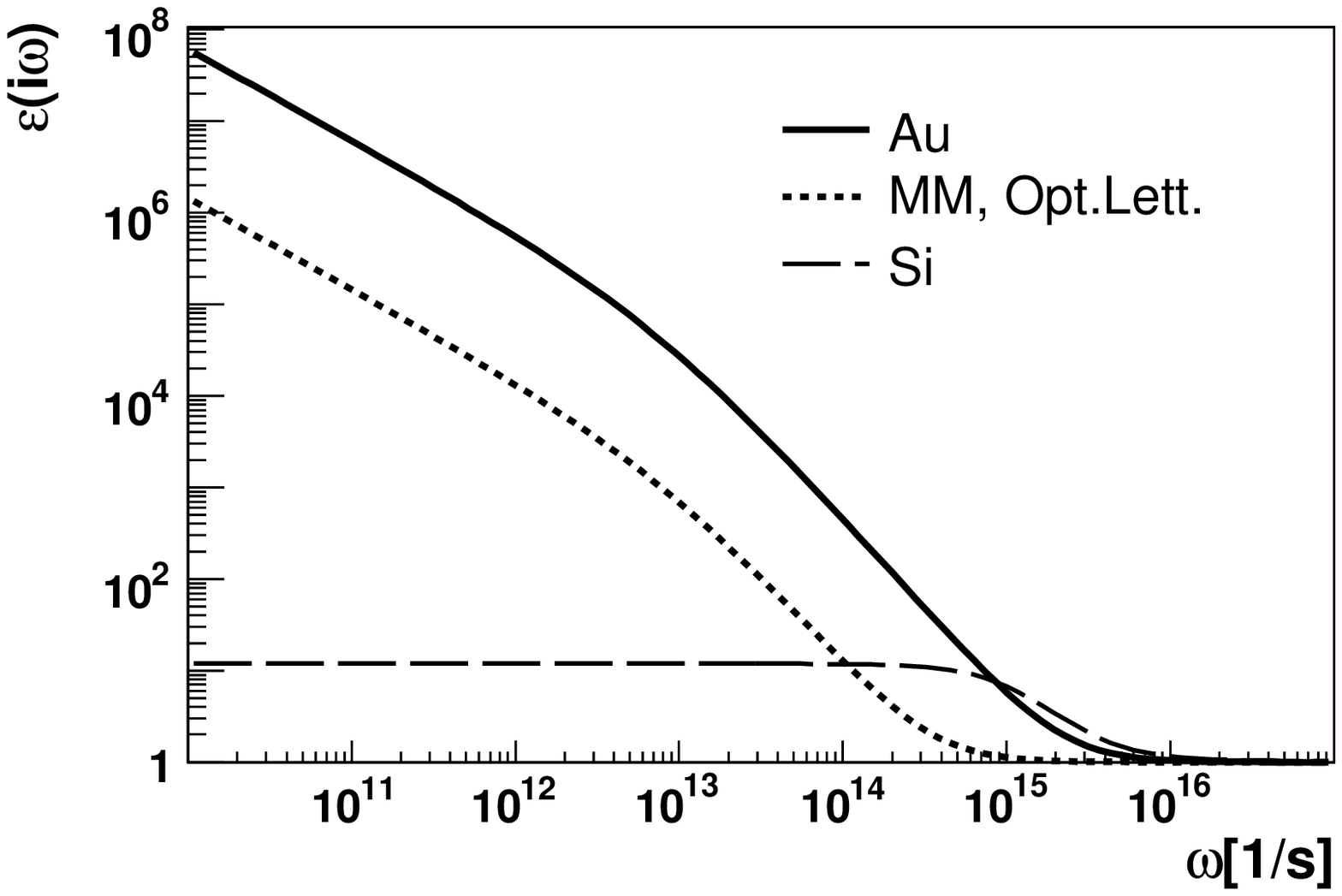,width=7cm}
\epsfig{file=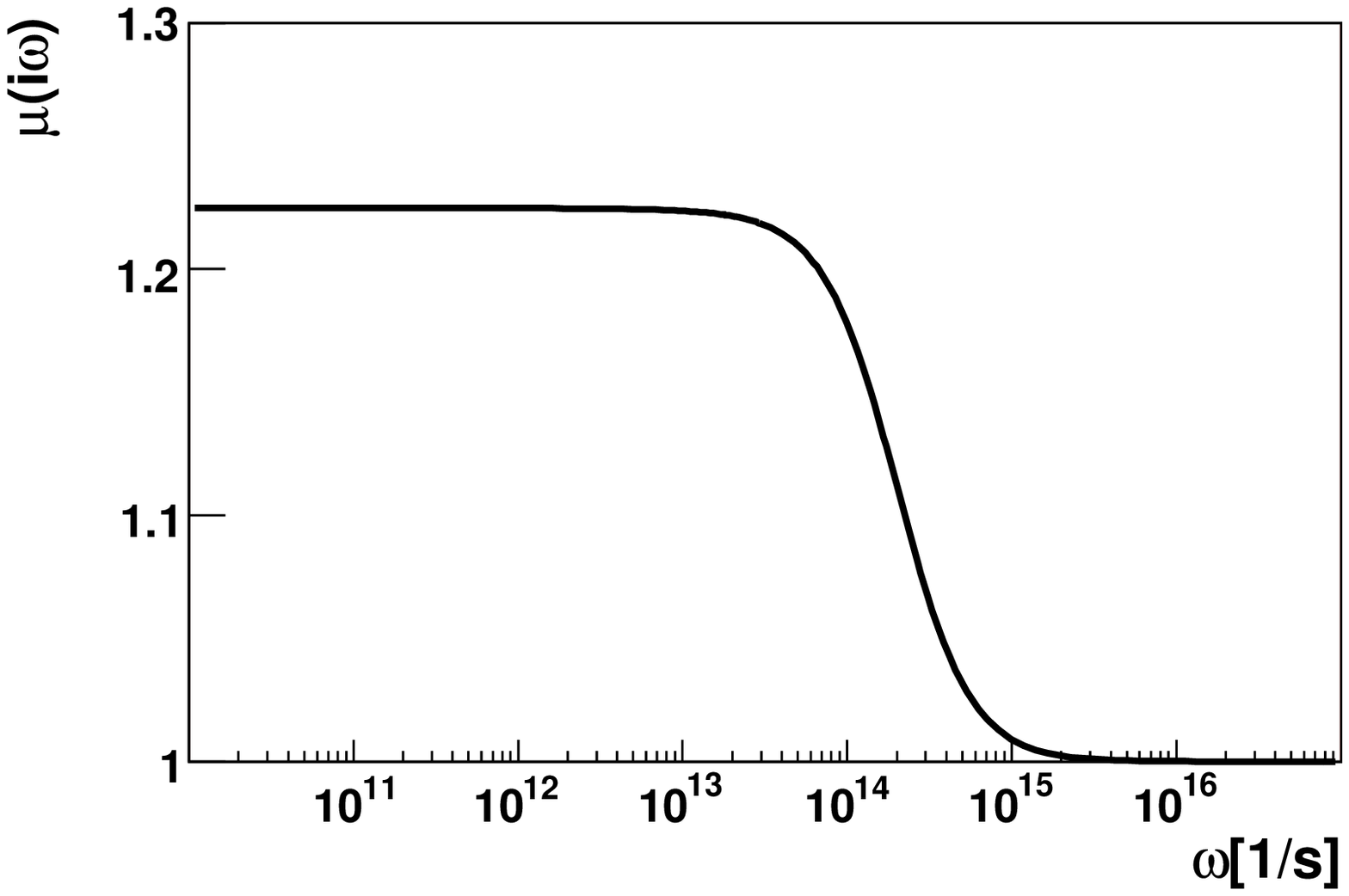,width=7cm}
\newline
\caption{ Left panel: $\varepsilon({\rm i}\omega)$ for
meta-material discussed in~\cite{OPT_LETT_1} in comparison with
gold and silicium.
Right panel: the magnetic permeability of the meta-material $\mu_(i\omega)$;
$\mu_{Au}=\mu_{Si}=1$.} \label{fig2}
\end{figure}
For the dielectric permittivity we have  taken the Drude response
with a small anti-resonance, the magnetic permeability has Lorentz form, Fig.~\ref{fig2},
$$
\varepsilon(\omega)=1-\frac{\omega_e^2}{\omega\,(\omega+{\rm i}
\Omega_e)}+\frac{\omega_{e1}^2}{\omega^2-\omega_0^2+{\rm i}\Omega_{e1}\omega}, 
\quad
\mu(\omega)=1-\frac{\omega_{m1}^2}{\omega^2-\omega_0^2+{\rm i}\Omega_{m1}\omega},
$$
with the position of the resonance defined by $\omega_0$.
Though the amplitude of the magnetic resonance is larger than the corresponding dielectric one,
the Drude-type term dominates in $\varepsilon(\omega)$. The magnetic
permeability of the meta-material is described by a
function which is characteristic for the dielectric permittivities
of the semiconductors, for example silicon~(Fig.\ref{fig2}).


The calculation of the Casimir force between golden mirror A and mirror B made of
this metamaterial was performed with the following values of the parameters:
 $ \omega_e/2\pi=3.6 \cdot
10^{5}$~GHz, \ $\Omega_e/2\pi = 8.9\cdot 10^{3}$~GHz, \
$\omega_0/2\pi=2.05\cdot 10^5$~GHz, \ $\omega_{e1}/2\pi =
2.04\cdot 10^4$~GHz, \  $\Omega_{e1}/2\pi =5.03\cdot
10^3$~GHz, \ $\omega_{m1}/2\pi= 9.72 \cdot 10^4$~GHz, \
$\Omega_{m1}/2\pi =1.1 \cdot10^4$~GHz. The
metamaterial being mainly dielectric, we predict no repulsion in the
present setup.

The
effective medium approach is valid for wavelengths longer than the
"lattice constant" of the meta-material.  To put it differently, the
theoretical estimations for the force are trustable  for plate
separations large in comparison with the "lattice constant" of the
meta-material. For more accurate results optical data in a wide
frequency range and for different incidence angles are needed.



\section{Conclusion}

When one of the mirrors is mainly magnetic and its magnetic permeability is described by the
Drude-Lorenz model the repulsion could be observed at the separations of the order $L\sim\lambda_0=2\pi c/\omega_0$.
However the force decreases rapidly with the distance.
That is why to get measurable Casimir  repulsion we  look for a material
with magnetic permeability $\mu\ne1$ at optical frequencies.

Non-magnetic conductive elements, such as
split ring resonators (SRR), being embedded into a dielectric lead to
nontrivial magnetic response of the compound~\cite{Shelby_Smith}.
A metamaterial with effective  permeability $\mu_{eff}$ is obtained when a lot of such elements are oriented
in different directions and positioned as
cubic lattice~\cite{IEEE_pendry}.
As the effective resonance frequency is defined by the geometric scale  $s$ of the resonator,
$\omega_0\sim1/\sqrt{LC}\sim1/s$~\cite{Tretyakov}, the latter
should be of several hundred nm size.
Though when the frequency of the field  approaches the plasma frequency
of the metal this estimation is not valid anymore, as the
electrons become insensitive to the variation of the field.
Experimentally it was shown that geometrical scaling
law for the resonance frequency brakes down at about 300 THz~\cite{klein}.

We seek for mainly magnetic material. The height of the magnetic resonance peak for a
single resonator increases with the filling fraction. On the contrary, the interaction
between the inclusions makes the resonance peak broader and lower. At the same time the
metallic inclusions change  the dielectric permittivity of the host media. It acquires
the properties of diluted metal or highly doped dielectric with the dielectric
permittivity dominated by the Drude term. In other words, providing us with needed
magnetic response the metallic structures add as well redundant dielectric permittivity,
that  makes the metamaterials  mainly dielectric. Consequently they  do not manifest
repulsion in the Casimir set-up. The metamaterials based entirely on dielectrics are more
appropriate. That could be, for example, arrays of dielectric spheres in a dielectric
matrix~\cite{Veselago2,Yannop} with the magnetic response arising from polaritonic
resonances.

\section{Acknowledgements}
The work was supported by European contract STRP 12142 NANOCASE.

\end{document}